# High-temperature THz dielectric spectra of $SrBi_2Ta_2O_9$ thick film

F. Kadlec, S. Kamba, P. Kužel, C. Kadlec, J. Kroupa

Institute of Physics, Academy of Sciences of the Czech Republic, Na Slovance 2,
182 21 Prague 8, Czech Republic

**Abstract**
Time-domain THz transmission experiment was performed on $SrBi_2Ta_2O_9$ thick film (deposited on sapphire substrate) at temperatures between 30 and 923 K and complex dielectric spectra of the film were determined. The lowest frequency optic phonon near 28 $cm^{-1}$ reveals slow monotonous decrease in frequency on heating with no significant anomaly near either the ferroelectric or ferroelastic phase transitions. An additional relaxation mode was resolved in the THz dielectric spectra at all temperatures and its slowing down to ferroelectric phase transition explains dielectric anomaly near $T_{c1}$. Second harmonic generation signal was observed in single crystal not only below $T_{c1}$ but also below $T_{c2}$, which confirms the presence of intermediate ferroelastic phase.

## 1. Introduction

Although ferroelectricity was discovered in $SrBi_2Ta_2O_9$ (SBT) already at the beginning of the 1960's [1,2], this material begun to be intensively investigated mainly in the last decade, because it has excellent polarization fatigue-free behaviour, low leakage currents and it can be prepared in extremely thin films without loss of bulk characteristics[3]. It allowed use of SBT films in commercial non-volatile ferroelectric memories. On other hand it is interesting that the structural phase transitions in SBT are not yet well characterized.

SBT crystallizes in so called Aurivillius structure, where the perovskite-type groups $[SrTa_2O_7]^{2-}$ and $[Bi_2O_2]^{2+}$ are stacked alternately along the pseudotetragonal $c$ axis[4, 5]. The $Bi_2O_2$ layers ad $TaO_6$ octahedra are considerably distorted and atomic displacements along the $a$ axis give rise to spontaneous polarization[6]. Ferroelectric structure has space group $A2_1am$ ($Z_{prim}$=2) and undergoes the phase transition (PT) to paraelectric phase at $T_{c1}$ near 600 K. Recently was discovered intermediate tetragonal ferroelastic phase with $Amam$ space group($Z_{prim}$ =2) [7, 8, 9] which undergous to paraelectric paraelastic phase with $I4/mmm$ ($Z_{prim}$ =1) structure above $T_{c2} \cong 770$ K. This second phase transition is improper ferroelastic (with doubling of the primitive unit-cell volume on cooling) and the lower temperature one is proper ferroelectric (no multiplication of the primitive unit-cell volume with respect to the ferroelastic phase)[9].

The dynamics of the PT's in SBT has been investigated first by Raman scattering [10, 11, 12], later by Brillouin scattering[13] and Fourier transform infrared (FTIR) reflectivity and transmission spectroscopy[9, 14, 15]. It was shown that the lowest frequency optic phonon near 28 $cm^{-1}$ (at 300 K) reduces its frequency on heating and even disappears from Raman spectra above $T_{c1}$ due to the change of selection rules. Therefore this mode was called optic soft mode. However, high-temperature FTIR transmission spectra revealed a weak monotonic softening down to 20 $cm^{-1}$ at 950 K with no significant anomaly near either $T_{c1}$ or $T_{c2}$. Such temperature behaviour cannot explain dielectric anomaly observed in low-frequency dielectric data near $T_{c1}$, therefore the name soft mode is not correct. We proposed existence of an additional relaxation mode, which will slow down to $T_{c1}$ and will cause dielectric anomaly near $T_{c1}$[9]. This relaxation we really observed in recent THz spectra below 300 K[14, 16]. The aim of this paper is the continuation of the previous work at higher temperatures and to bring THz dielectric spectra of SBT thick film (same like in Ref. [14]) up to 923 K, which will show the high temperature behaviour of the relaxation and of the low-frequency phonons. This will allow us to discuss the mechanism of the phase transitions in SBT.

## 2. Experimental

Thick SBT film with thickness 5.5 μm was prepared by pulsed laser deposition on (0001) oriented sapphire by A. Garg and Z.H. Barber in Cambridge.
THz
Commercial high-temperature cell SPECAC P/N 5850 with sapphire windows was specially adapted with second cooling circle for THz setup.
S
Second harmonic generation (SHG) signal was measured on SBT single crystal with dimensions of about 2 mm x 1mm x 0.1 mm, its large face was perpendicular to the *c* axis. The same crystal was used in a previous studies [9, 14]. Q-switched Nd-YAG laser served as a light source. The pulse energy was 0.2 mJ and the beam was only slightly focused (2w~300 μm) on the sample. After filtering the fundamental frequency the SHG signal was detected by photomultiplier by a boxcar integrator (PAR 162). The sample was placed in the custom made heating cell and temperature was continuously swept with rate of 5 K/min up to 900 K. It was possible to monitor the position of the beam on the sample in the oven.

## 3. Results and discussion

The complex dielectric spectra of SBT thick film are shown at various temperatures up to 923 K in figure 1. The spectra were fitted with a sum of two damped Lorentz oscillators (describing the polar phonons) and a Debye relaxation accounting for the central mode (CM):

$$\varepsilon^*(\omega) = \varepsilon'(\omega) + i\varepsilon''(\omega) = \frac{\Delta\varepsilon_r \omega_r}{\omega_r + i\omega} + \sum_{j=1}^{2} \frac{\Delta\varepsilon_j \omega_j^2}{\omega_j^2 - \omega^2 + i\omega\gamma_j} + \varepsilon_\infty. \quad (1)$$

$\omega_r$ and $\Delta\varepsilon_r$ are the relaxation frequency and dielectric strength of the CM, respectively. $\omega_j$, $\gamma_j$, and $\Delta\varepsilon_j$ denote the eigenfrequencies, damping, and contribution to the static permittivity from the j-th polar phonon mode, respectively, and $\varepsilon_\infty$ describes the high-frequency permittivity originating from the electronic polarization and from polar phonons above the spectral range studied by TDTS. Typical example of the experimental dielectric spectrum at 573 K together with the fitted curves is depicted in figure 2. Two kinds of the fits were used; with and without relaxational CM. The model without CM satisfactorily describes $\varepsilon^*(\omega)$ above 15 cm$^{-1}$, however does not describe the low-frequency part of the $\varepsilon^*(\omega)$ spectra and gives also too low static permittivity ($\varepsilon'$=130 near $T_{c1}$). Although the accuracy of our data is below 10 cm$^{-1}$ limited, the fit with CM describes the experimental spectra qualitatively much better. Temperature dependences of the eigenfrequencies of the CM and of the lowest frequency phonon together with their dielectric strengths are depicted in figure 3. The CM exhibits slowing down to $T_{c1}$ and again increase at higher temperatures. The strength of the CM f = $\Delta\varepsilon_r\omega_r$ was taken temperature independent, therefore the maximum in $\Delta\varepsilon_r$ (as well as in static permittivity $\varepsilon_0$) was obtained at $T_{c1}$. The first phonon exhibits no anomaly with temperature, only slow decrease of its eigenfrequency $\omega_1$ and increase of the dielectric strengths $\Delta\varepsilon_1$ was observed on heating. So only the CM can explain dielectric anomaly near the ferroelectric PT at $T_{c1}$.

Note that the value of permittivity maximum $\varepsilon'_{max}$ at $T_{c1}$ is different by various authors and oscillates between 280 [1] and 580 [2] in stoichiometric ceramics, in the case of nonstoichiometric $Sr_{1\pm x}Bi_{2\pm y}Ta_2O_9$ ceramics can be $\varepsilon'_{max}$ even higher, but $T_{c1}$ also raises [17]. On other hand it is known that $\varepsilon'$ in SBT is size dependent only in the films with thickness less than 200 nm, which is much less than the 5.5 μm we used. In spite of it static permittivity obtained from our TDTS experiment is slightly lower than the value published in ceramics. It can be explained by cracks [18] and fine scratches which we observed in the film with microscope. Our TDTS is sensitive on permittivity in the plane of the film and the cracks with zero permittivity can strongly reduce the dielectric response [18]. Unfortunately, we were not able to check directly the low frequency permittivity of our film in the sample plane, because it needs interdigital electrode method which is not accessible in our lab. Room-temperature THz spectra of the SBT film were measured many times during the last two years and dielectric response in THz range decreased with time (compare e.g. figure 1 in Ref. [15] and figure 1 in this article) just due to the increasing concentration of scratches in our film from tweezers (confirmed in microscope).

Sample age had no influence on phonon frequencies. Figure 3 shows temperature dependence of $\omega_1$ frequency obtained from TDTS and FIR transmission [15] spectra, the same values and temperature dependences were observed. FIR transmission spectra were fitted without CM, therefore also its dielectric strength $\Delta\varepsilon_1$ is higher than $\Delta\varepsilon_1$ obtained from TDTS with CM. Oscillator fits of THz spectra without CM give the same $\Delta\varepsilon_1$ parameters like the fit in Ref. [15], so different values of $\Delta\varepsilon_1$ in figure 3 are not due to the different experimental methods but due to the different fitting procedure. Let us stress that the lowest frequency phonon is not the optic soft mode responsible for the ferroelectric PT, because it exhibits no anomaly in $\omega_1$ frequency near either $T_{c1}$ or $T_{c2}$.

Stachiotti et al. [19] studied lattice dynamics in SBT using the frozen-phonon approach and proposed soft mode of $E_u$ symmetry (assignment in paraelectric *I4/mmm* structure), which involves Bi atoms vibration relative to the $TaO_6$ perovskite-like blocks. However, the freezing of the polar $E_u$ mode does not lead to the observed ferroelectric phase of $A2_1am$ symmetry as well as to intermediated ferroelastic phase. Perez-Mato et al. [20] studied the stability of phases in SBT by means of ab-initio calculations and came to the conclusion that two unstable modes are responsible for the two subsequent phase transitions. The stronger instability concerns the $X_3^-$ nonpolar mode (this mode from X point of Brillouin zone involves antiphase octahedra $TaO_6$ tilting), whose freezing yields to Amam intermediate phase with doubled unit cell. Subsequent condensation of the $E_u$ polar mode is responsible for the $A2_1am$ ferroelectric phase. From our data follows that the $E_u$ mode is not phonon but relaxation mode describing probable an anharmonic vibration of Bi ions at the Sr sites (anti-site defects are known in SBT [21]). The $X_3^-$ mode can in principle become IR and/or Raman active below $T_{c2}$, however this mode was not observed in our THz spectra up to 80 cm$^{-1}$ as well as in our previous FIR transmission spectra up to 140 cm$^{-1}$. It means that this mode is either too weak or lies above our frequency limit. FIR reflectivity and Raman scattering spectra were taken in the broad spectral range of all possible phonons, but the spectra were taken never above $T_{c2}$, so no conclusion about frequency of $X_3^-$ mode below $T_{c2}$ can be done.

Figure 4 shows temperature dependence of the SHG signal from SBT single crystal. Large SHG signal is seen in ferroelectric phase below $T_{c1}$. Two order of magnitude lower signal, but still one order of magnitude higher than our detection limit, remains between $T_{c1}$ and $T_{c2}$. It gives evidence about the loss of center of symmetry in ferroelastic phase. Above $T_{c2}$ is SHG signal zero within accuracy of our experiment. The SHG signal in ferroelastic phase was detected for parallel polarization, surprisingly no SHG was observed for the perpendicular polarization. Its intensity was probably below our noise limit. However, the SHG signal was similar for both polarizations in ferroelectric phase below $T_{c1}$.

**4. Conclusion**

**Acknowledgments**

**Figure captions**

**Figure 1** Experimental real and imaginary part of complex THz permittivity in SBT plotted at various temperatures.

**Figure 2** Experimental THz dielectric spectra at 573 K together with different fits (see the text).

**Figure 3** a) Temperature dependences of frequencies of CM $\omega_r$ and of the first optic phonon $\omega_1$. b) Temperature dependences of dielectric strengths of the CM ($\Delta\varepsilon_r$) and of the lowest frequency phonon ($\Delta\varepsilon_1$). Open squares mark $\Delta\varepsilon_r$ obtained from the fit with CM, stars come from oscillator fit without CM. Temperature dependence of static permittivity $\varepsilon_0$ is also shown.

**Figure 4** Temperature dependence of second harmonic generation signal in SBT single crystal for parallel polarizations. The noise limit is marked by dashed line.

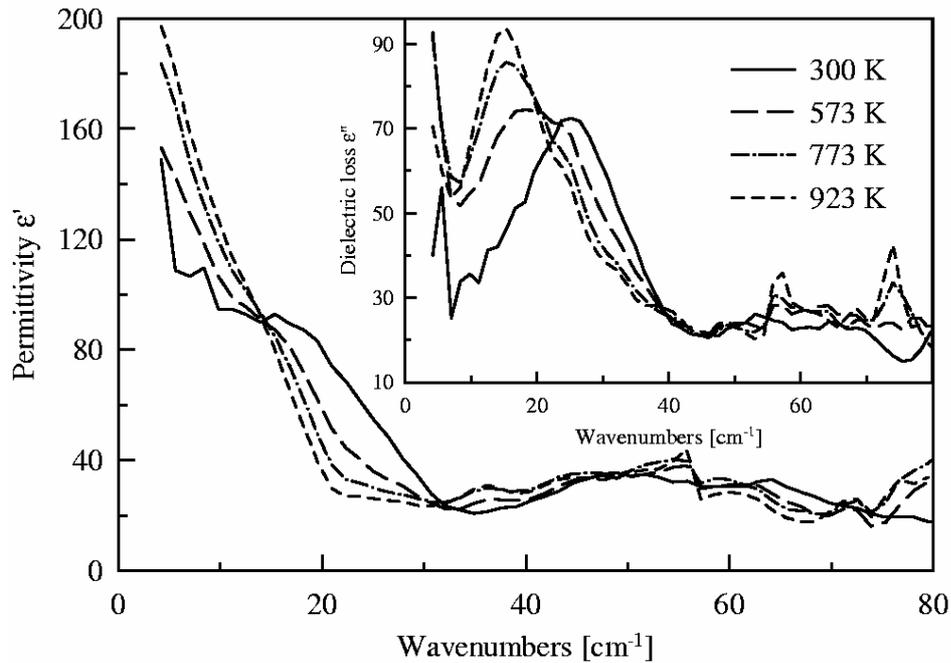

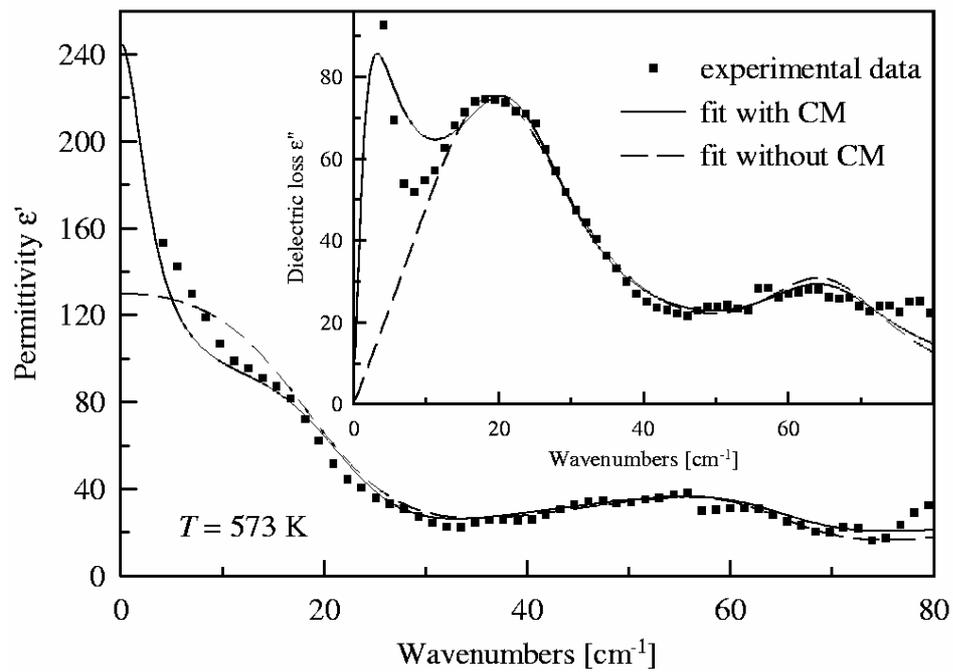
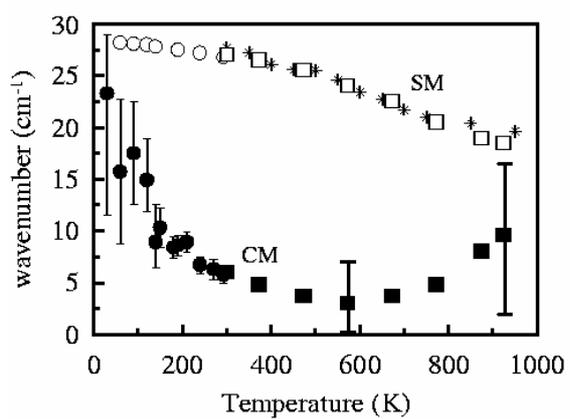
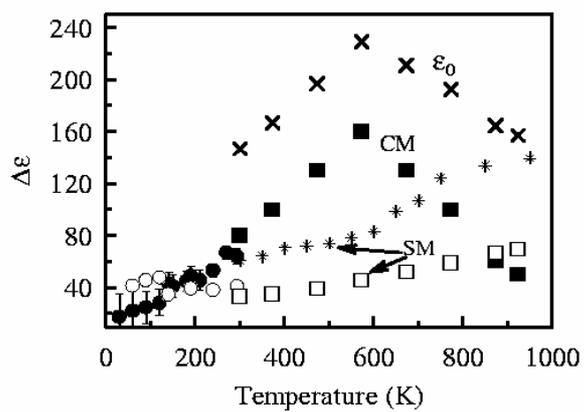

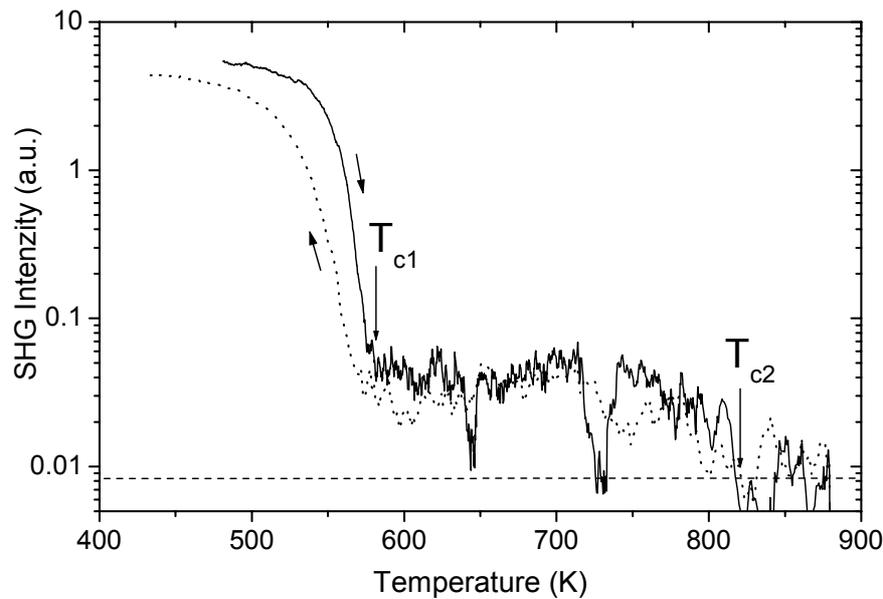